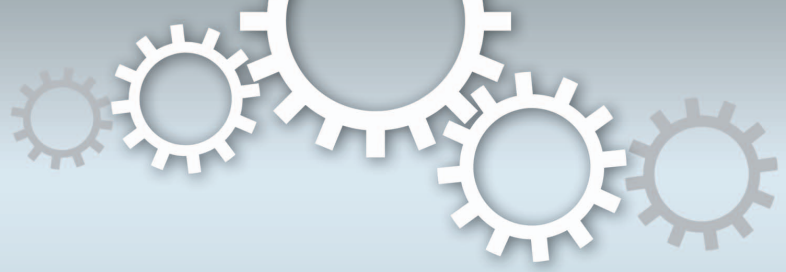

Correspondence and
requests for materials
should be addressed to
M.V.T. (mtomasello@
ethz.ch)

# The role of endogenous and exogenous mechanisms in the formation of R&D networks

Mario V. Tomasello[1], Nicola Perra[2], Claudio J. Tessone[1], Márton Karsai[3] & Frank Schweitzer[1]

[1]Chair of Systems Design, Department of Management, Technology and Economics (D-MTEC), ETH Zurich, Weinbergstrasse 56/58, 8092 Zurich, Switzerland, [2]Laboratory for the Modeling of Biological and Socio-technical Systems, Northeastern University, Boston, MA 02115, USA, [3]Laboratoire de l'Informatique du Parallélisme, INRIA-UMR 5668, IXXI, ENS de Lyon, 69364 Lyon, France.

We develop an agent-based model of strategic link formation in Research and Development (R&D) networks. Empirical evidence has shown that the growth of these networks is driven by mechanisms which are both endogenous to the system (that is, depending on existing alliances patterns) and exogenous (that is, driven by an exploratory search for newcomer firms). Extant research to date has not investigated both mechanisms simultaneously in a comparative manner. To overcome this limitation, we develop a general modeling framework to shed light on the relative importance of these two mechanisms. We test our model against a comprehensive dataset, listing cross-country and cross-sectoral R&D alliances from 1984 to 2009. Our results show that by fitting only three macroscopic properties of the network topology, this framework is able to reproduce a number of micro-level measures, including the distributions of degree, local clustering, path length and component size, and the emergence of network clusters. Furthermore, by estimating the link probabilities towards newcomers and established firms from the data, we find that endogenous mechanisms are predominant over the exogenous ones in the network formation, thus quantifying the importance of existing structures in selecting partner firms.

T he increasing importance of Research and Development (R&D) activities has spurred the formation of partnerships between firms and other economic actors, whose number has significantly raised over the last four decades[1]. Indeed, inter-firm alliances bring a certain number of advantages, such as reputational effects, technological risk sharing and resource pooling. For these reasons, they have become an important part of many firms' strategy, especially in sectors characterized by high technological dynamism and uncertainty[2].

R&D alliances can be represented as networks. Nodes describe *firms* and links their *R&D alliances*. A number of empirical works have characterized the properties of these networks[2–7]. Furthermore, other theoretical studies have tried to capture, predict and model some features of R&D networks, such as their topology or profit efficiency[8–12].

In this context, two types of mechanisms have been proven crucial in the formation of new R&D alliances[9]: *endogenous* mechanisms (previous alliances and previous network structures) and *exogenous* mechanisms (exploratory search for new partners). However, both empirical and theoretical studies have mainly focused only on one of the two mechanisms, also called "network endogeneity"[7,13–15] and "exogenous partner selection"[16–18] respectively. Typically, the concept of endogenous and exogenous mechanisms has been used in the management literature with respect to the belonging of the firms to the R&D network. We follow such definition and refer to an alliance involving a partner that is already part of the R&D network as "endogenous". Likewise, an alliance involving a partner that is not part of the R&D network yet is referred to as "exogenous". While the endogenous mechanisms depend on the firms' social capital (describing their position in the network), the exogenous mechanisms are affected by the firms' technological and commercial capital. A firm's social capital can be further explained by two variables[19,20]: its *prominence* – i.e. the history of its previous alliances – and its *cohesiveness*, defined as the set of its direct and indirect links with other firms in the network. In this regard, some empirical studies[9,14] found that several firm "clusters" populate the R&D network, thus giving rise to different kinds of alliances depending on the firms' position in the network. In particular, three categories of R&D alliances have been identified: i. within-cluster alliances (the partners belong to the same cluster); ii. semi-distant alliances (the partners form a so-called "shortcut" between two different clusters); iii. distant alliances (at least one of the partners is an isolated node, i.e. a newcomer firm). Obviously, a certain number of R&D alliances is not explained by the partners' social capital – think, for instance, of alliances involving start-up companies or financial institutions that have no previous experience in R&D activities. One rationale for the search of this kind of partners, whose technological and commercial capital plays a crucial role, is that they can provide access to new information or unique technical knowledge. However, neither the network endogeneity nor the exogenous partner selection, taken independently, are able to explain the topology of observed R&D networks. Endogenous mechanisms alone would lead to more and more centralized network structures over time, which we do not observe in reality[6]. On the other hand, exogenous mechanisms alone would lead to more regular networks topologies, which we do not observe neither.

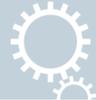

Inspired by these observations, here we introduce an agent-based model that includes and allows to modulate the weight of both endogenous and exogenous mechanisms for alliance formation. The model aims at reproducing the main global properties and a set of microscopic measures (including degree, local clustering and path length distributions) of real R&D networks. To this purpose, we test the model against one of the most comprehensive R&D alliance dataset available nowadays, consisting of a time-stamped list of global R&D alliances in the period 1984–2009. The validation of the model and the tuning of its parameters give insights into the micro-level decisions operated by the agents and, consequently, the growth of the network itself. In particular, the results obtained indicate that endogenous mechanisms play a more relevant role in the network growth than the exogenous ones.

## Results

**Empirical evidence.** We built our empirical R&D network using the *SDC Platinum* database[21] that reports approximately 672,000 publicly announced alliances in all countries, from 1984 to 2009, with a granularity of 1 day, between several kinds of economic actors (including manufacturing firms, investors, banks and universities). We then select all the alliances characterized by the "R&D" flag; after applying this filter, a total of 14,829 alliances, connecting 14,561 firms, are listed in the dataset. An R&D alliance is defined as an event of partnership between two or more firms, that can span from formal joint ventures to more informal research agreements, specifically aimed at research and development purposes. Every firm listed in the dataset is associated with a SIC (Standard Industrial Classification) code – a US-government system that allows us to univocally assign each firm to its corresponding industrial sector. We employ this sectoral classification to test the invariance and robustness of some empirical properties of the R&D network across sectors. A salient feature of the R&D alliances in the SDC dataset is the variable number of partners they involve. Most of the collaborations (93%) are stipulated between two partners, but some alliances – the so-called *consortia* – involve three or more partners. The distribution of the number of firms per alliance event, as shown in Fig. 1, spans one order of magnitude and is right-skewed. This feature holds independently of the industrial sector to which the alliance partners belong (see the Supplementary Information, SI).

In our network representation, we draw an undirected link connecting two nodes every time an alliance between the two corresponding firms is announced in the dataset. When an alliance involves more than two firms, we assume that all the corresponding nodes are connected in pairs, forming a fully connected clique. This choice derives from the fact that consortia, although representing only a minority of the alliances, require great coordination and resource availability from the partners. More precisely, following this procedure we obtain a total of 21,572 links from the 14,829 alliance events listed in the dataset. However, in the definition of our model, we do not make any difference between a consortium and a "standard" two-partner alliance, which is only a special case of it (and can be thought of as a fully connected clique of size 2).

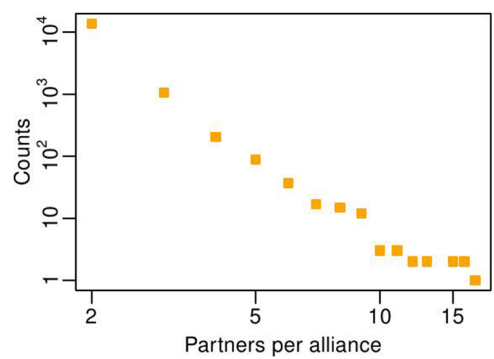

**Figure 1 | Distribution of the number of partners per alliance, as measured from the SDC alliance dataset.**

Another distinctive measure we introduce and analyze in this study is the firms' *activity* distribution[22]. Developed in the field of temporal networks[23], the activity has been studied on various datasets, such as coauthorship, online microblogging, or actor/movie networks. Applying this concept to an inter-organizational R&D network is a logical consequence of these recent developments. To the best of our knowledge, no previous work has measured this quantity on a set of real firms involved in R&D alliances by using empirical data. We define the empirical *activity* $a_{i,t}^{\Delta t}$ of a firm $i$ at time $t$, over a time window $\Delta t$, as the number of alliance events $e_{i,t}^{\Delta t}$ involving firm $i$ in the time window $\Delta t$ ending at time $t$, divided by the total number of alliance events $E_t^{\Delta t}$ involving *any* firm in the same time period:

$$a_{i,t}^{\Delta t} = \frac{e_{i,t}^{\Delta t}}{E_t^{\Delta t}}. \tag{1}$$

The activity expresses the probability that a firm takes part in any alliance event occurring in a given time window. We test four time window lengths $\Delta t$ equal to 1, 5, 10 and 26 years and we find that the empirical firm activity distribution is virtually independent of the chosen $\Delta t$. We report our finding in Fig. 2. The firm activity distributions are right skewed and dispersed over several orders of magnitude, as in many other social and technological systems[24–26]. Contrary to most of the R&D network indicators, that display strong variability and dependence on time[6], the activity is a stable attribute that can be assigned to firms and effectively estimate their prom-inence. Indeed, empirical firm activities are robust with respect to the time $t$ at which they are measured: shifting the time window – of any length $\Delta t$ – along the 26 years reported in the dataset does not affect the results (see SI). In addition, we find that the activity distribution is robust to the sectoral classification of the firms (see SI).

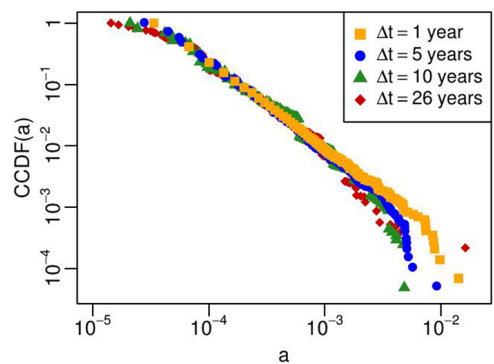

**Figure 2 | Complementary cumulative distribution function (CCDF) of the empirical firm activities, measured on the SDC dataset with 4 different time windows $\Delta t$ of 1, 5, 10 and 26 years.** When the time window is shorter than 26 years (the entire dataset observation period), we compute the activity by shifting the time window in 1-year increments and then we average the results.

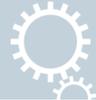

**Table 1 | Observed values of average degree $\langle k \rangle^{OBS}$, average path length $\langle l \rangle^{OBS}$ and global clustering coefficient $C^{OBS}$ for the empirical R&D network**

| Measure | Value |
|---|---|
| $\langle k \rangle^{OBS}$ | 2.736 |
| $\langle l \rangle^{OBS}$ | 5.412 |
| $C^{OBS}$ | 0.101 |

The empirical R&D network, as well as the networks we generate by means computer simulations, are then characterized with respect to the three following measures:

- *Degree*. We define the degree $k_i$ of a node $i$ as the number of its neighbors. In reality, two firms can have more than one alliance on different projects. Nevertheless, as we aim at studying the connections between firms, and not the number of alliances a firm is involved in, we discard this information and use *unweighted* links in our network representation. Finally, we call $\langle k \rangle$ the average of this measure over all nodes in the network.
- *Path length*. The path length $l_{ij}$ between two nodes $i$ and $j$ is defined as the lowest number of links that must be traversed in order to reach $j$ from $i$ or viceversa ($l_{ij} = l_{ji}$ in an undirected network such as the R&D network). We then define $\langle l \rangle$ as the average of this measure over all pairs of nodes in the network. If the nodes $i$ and $j$ belong to disconnected components of the network, we simply discard them for the computation of the average path length.
- *Clustering coefficient*. The local clustering coefficient $c_i$ of a node $i$ measures the extent to which its partners are connected to each other in their turn. It is defined as the ratio of the existing links between the neighbors of the node $i$ to all possible links between these neighbors[27]. We also characterize the network with respect to its global clustering coefficient $C$, also known as transitivity. It is defined as the ratio of closed triads (i.e. groups of three nodes mutually linked) to all paths of length two in the network. It should be noted that the global clustering coefficient $C$ is not equal to the average $\langle c \rangle$ of the local clustering coefficients over all nodes in the network[27], although some studies use the latter measure. We prefer to use $C$ because it has a direct interpretation and – differently from $\langle c \rangle$ – is not affected by the large number of low-degree nodes that populate the network.

We report in Table 1 the observed values of average degree $\langle k \rangle^{OBS}$, average path length $\langle l \rangle^{OBS}$ and global clustering coefficient $C^{OBS}$ for the empirical R&D network. The values are related to the aggregate network, obtained by considering all the alliances occurring during our observation period 1984–2009; from now on, we refer to the aggregate R&D network simply as the empirical R&D network.

This set of measures gives us an intuition of the global properties of the network. In particular, the average degree indicates a quite sparse topology, i.e. firms have on average only 2.7 partners. Nevertheless, the global clustering coefficient indicates a moderate level of transitivity, and the average path length clearly reveals the small-world properties of the R&D network. In order to deepen our empirical exploration, we report a set of microscopic measures, namely the distributions of node degrees, path lengths, local clustering coefficients and network component sizes. A *component* of the network is defined as a set of nodes which are connected to each other by at least one path; the number of these nodes is the size of the component. As we report in Fig. 3, the degree distribution is heavy-tailed, indicating heterogeneities in the connectivity patterns. We will later show that, despite these properties of the degree distribution, the average degree is sufficient for us to reproduce many microscopic properties of the network. The average path length distribution is instead peaked around the mean value 5, whilst the local clustering coefficient is quite spread in its possible range, i.e. [0, 1], indicating large heterogeneities in local transitivity. The component size distribution shows the presence of a giant component in the R&D network (containing roughly 60% of the nodes), plus a myriad of smaller components down to size 2.

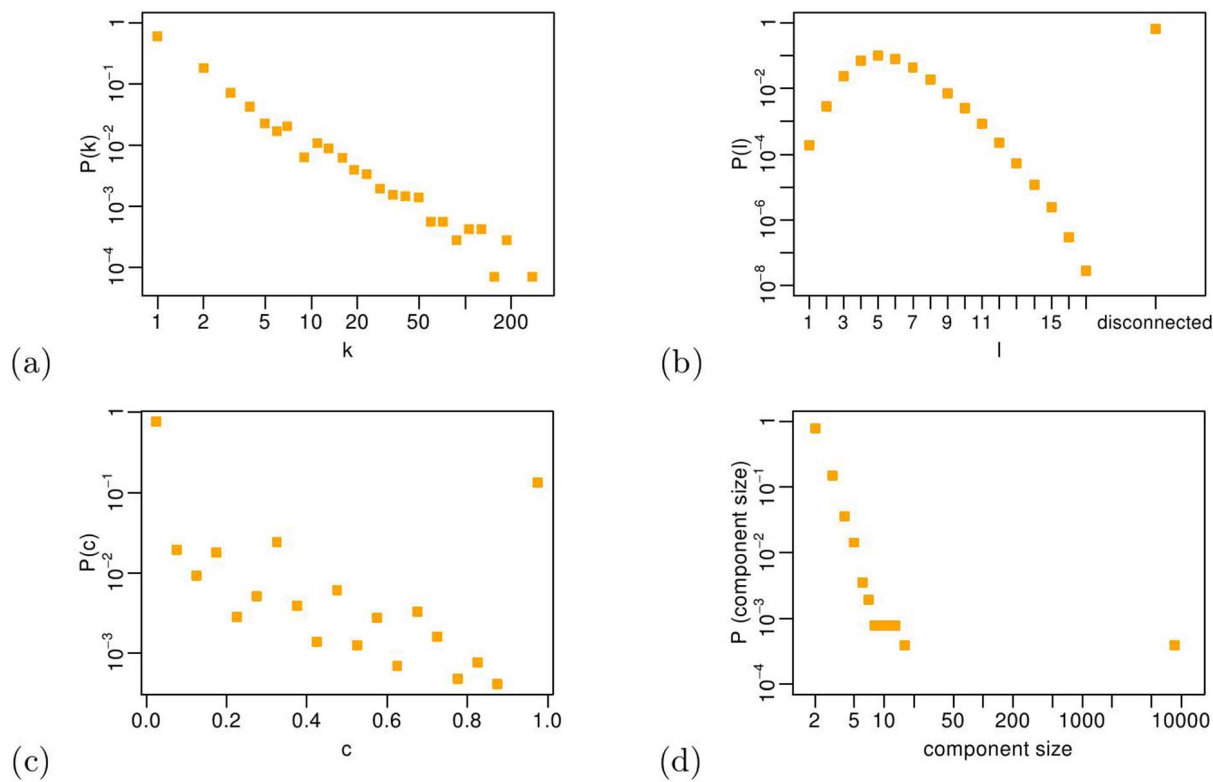

**Figure 3 | Distributions of node degrees (a), path lengths (b), local clustering coefficients (c) and component sizes (d) for the real R&D network.**

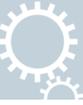

Finally, we turn our attention to the modular properties of the R&D network. It has been acknowledged that networks, in many different domains, are organized in modules or clusters, characterized by groups of tightly connected nodes[28,29]. R&D networks are not an exception[6]. Interestingly, the formation of such clusters is not explained by factors like the firms' industrial sectors or their geographical distribution[9]. Indeed, firms belonging to different sectors and located in different countries can populate the same network cluster. However, clusters in R&D networks have never been theoretically defined; they have been only empirically detected by means of simple K-means algorithms and used to obtain rough indications about the inter-firm alliance activity[9]. We perform a community detection on the empirical R&D network by using the Infomap algorithm[30] and report our findings in Fig. 4. We detect the presence of approximately 3,500 clusters in the R&D network, whose size distribution appears to be dispersed and right skewed, displaying a maximum cluster size of about 200 firms and a minimum cluster size of 2. In Fig. 4 we also provide a representation of the empirical R&D network; for the sake of visualization, we consider only the 30 largest firm clusters and depict them by grouping the corresponding nodes in 30 distinct regions of the plot area. Finally, we compute the so called modularity score of the empirical R&D network, to quantify the goodness of such division of the network in clusters. More precisely, we use a normalized version $Q$ of the modularity coefficient[27], defined such that $Q = 1$ in case of a perfectly modular network, where links are formed only within the same cluster. Likewise, $Q = -1$ for a perfectly anti-modular network, where links connect only nodes belonging to distinct clusters, and $Q = 0$ for a network where links are formed at random. We do not report here the complete definition of the normalized modularity coefficient $Q$, because outside of the scope of this work.

For the empirical R&D network, we observe a value of $Q^{OBS} = 0.68$, remarkably high if compared to other examples of real networks[31]. To check whether such a high value is indicative of a real modular structure in the R&D network, and not only an artifact caused by its size and density[32], we compute the modularity scores $Q$ for a set of 500 randomly generated networks having the same degree sequence as the empirical one. We find that the observed value $Q^{OBS} = 0.68$ is significantly greater (with a $p$-value computationally indistinguishable from zero) than the $Q$ scores obtained for the 500 randomly generated networks, which are normally distributed around a mean value of 0.570, with an extremely small variance of 0.001. Such a result constitutes a significative indicator of modularity in the R&D network.

**The model.** The empirical observations of the R&D network indicate clear heterogeneities in activity and connectivity patterns, small-world features, a moderate level of transitivity, and a highly modular structure, not simply associated to the firms' industrial sectors. Starting from this evidence, here we present a novel modeling framework for R&D networks. We consider a network composed of $N$ nodes; each of them is endowed with two fundamental attributes, an *activity* and a *label*. Such attributes define the nodes' interaction rules, which are organized in five distinct phases, as described below:

**Node activation.** We assign to each of the $i = 1, \ldots, N$ nodes an activity $a_i$, analogous to the empirical activities extracted from the SDC dataset. Indeed, we sample without replacement all the values $a_i$ from the empirical activity distribution. The activities we assign to the $N$ nodes are computed by considering the entire observation period 1984–2009 (therefore $a_i \equiv a_{i,t=2009}^{\Delta t = 26\,\text{years}}$). Given the strong robustness of empirical activities to the time window, we decide to use the 26-year window, because it contains complete information about the dataset and gives activities $a_i$ that are always strictly greater than 0 (all firms listed in the SDC dataset, by definition, must be involved in at least 1 alliance). The activity defines the propensity of each node to be involved in an R&D alliance event. We use this quantity to model the activation probability of each firm. In particular, at every time step, a node $i$ initiates an alliance with probability $p_i = \eta a_i \Delta t$, and the number of active nodes $N_A$ is

$$N_A = \eta \langle a \rangle N \Delta t, \qquad (2)$$

where $\langle a \rangle$ is the average node activity and $\eta$ is a rescaling factor that allows to adjust the activation rates, and consequently the number of active nodes per time step. We find that the model is strongly robust to the choice of $\eta$, showing no measurable changes for $\eta$ ranging from $10^{-5}$ to 1; however, we fix $\eta = 0.01$ to obtain $N_A$ roughly equal to 1.5, the number of alliance events per day actually reported in the dataset. Without loss of generality, we fix $\Delta t = 1$.

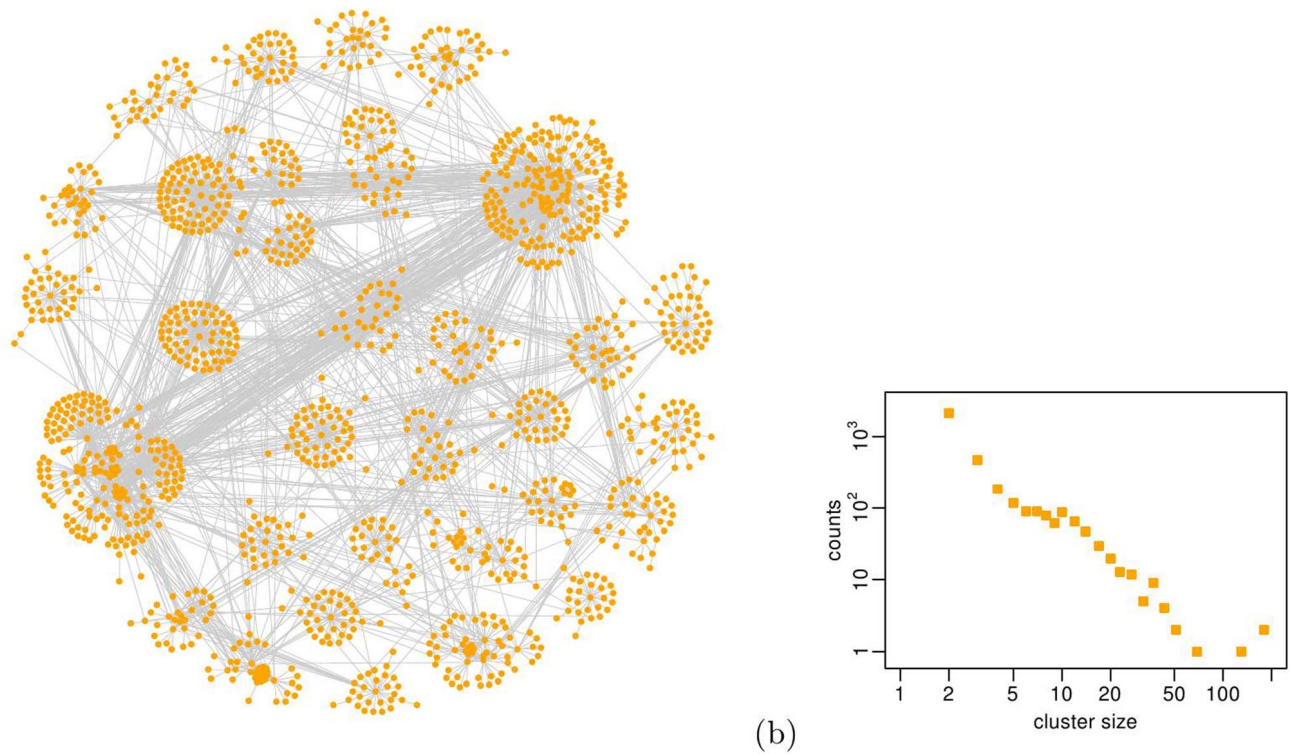

**Figure 4** | (a) Visual representation of the empirical R&D network (we use the Fruchterman-Reingold layout[42]), considering only the 30 largest clusters detected by the Infomap algorithm. Distinct clusters are represented by grouping nodes in distinct regions of the plot area. (b) Size distribution of the clusters in the empirical R&D network.

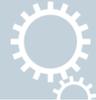

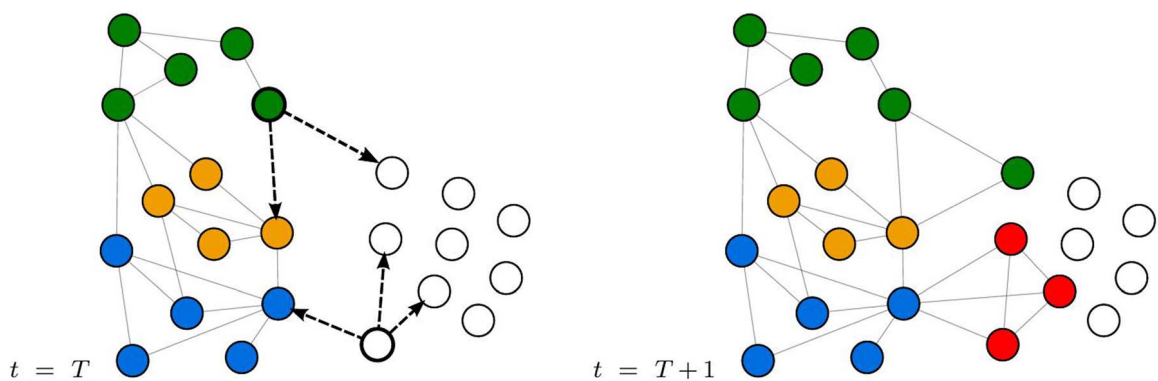

**Figure 5 | Two representative examples of label propagation.** A labeled node (whose label is depicted in green) chooses to form an alliance with $m = 2$ partners, one having a different label (depicted in yellow) and one non-labeled, at time $t = T$. The initiator propagates its green label at time $t = T + 1$ only to the previously non-labeled node. The link with the yellow node is still formed, but the label propagation does not occur. Likewise, a non-labeled node gets activated at time $t = T$ and forms an alliance with $m = 3$ partners, two non-labeled nodes and one labeled (blue) node. The non-labeled initiator takes a new arbitrary label (depicted in red) at time $t = T + 1$ and propagates it only to its previously non-labeled partners. The red label is not propagated to the blue node, even though the links are regularly formed.

**Selection of the alliance size.** When a node gets activated, it selects the number of partners $m$ with whom the alliance is formed. We assume that the value of $m$ is totally independent of any characteristic of the active node: we sample it, without replacement, from the empirical distribution of number of partners per alliance. In other words, we shuffle the sequence of number of partners per alliance (directly measured from the dataset) and then extract a value every time an activation event occurs; $m$ can be thought of as the number of partners involved in every alliance event, diminished by 1, because the active node is not counted twice.

**Label propagation.** As shown in the previous section, the real R&D network is organized in clusters of tightly interconnected nodes. However, these clusters are not isolated; previous studies[6,9] have detected the existence of "shortcuts" connecting different clusters, as well as the formation of alliances with new partners not yet belonging to the R&D network. This observation suggests that firms diversify some of their alliances, rather than just establishing collaborations within a specific cluster. We model this feature assuming that each of the $N$ nodes is endowed with an attribute named *label*. This attribute is unique – i.e. every node can have only one label at any time – and fixed – once a node assumes a label, it does not change –. Labels model the belonging of the firms to different groups that they implicitly define with their shared practices and commonly recognized behaviors: in other words, a label symbolizes the membership of the firm in a well defined and recognized "club" or "circle of influence". In addition, we assume that such membership can be transferred to other firms as a consequence of an alliance, provided that they are not part of any circle of influence yet. In our network representation, every alliance initiator does indeed propagate its label to all of its $m$ partners, if they are non-labeled. At the beginning of every simulation, all nodes are *non-labeled*, meaning that their membership attribute is blank. There are two ways a non-labeled node can assume its label: (i) the node either receives the label from another node, if the latter initiates an alliance, or (ii) it takes an arbitrary and unique label when it becomes active for the first time (see Fig. 5).

**Selection of the partner categories.** The presence of node labels induces different types of alliances, that we explicitly distinguish in our model (see Fig. 6). In particular, if the initiator is a labeled node, it represents an *incumbent* firm, i.e. a firm that has already been involved in at least one alliance. In this case, the initiator can link to a labeled node having the same label (with probability $p_s^L$), or to a node having a different label ($p_d^L$), or to a node without label ($p_n^L$). If the initiator is a non-labeled node, it represents a *newcomer* firm, i.e. a firm that has not been involved in any alliance event. In this case, the initiator can link to a labeled node (with probability $p_l^{NL}$), or to another non-labeled node ($p_{nl}^{NL}$). The five probabilities associated to these occurrences, represented in Fig. 6, are the free parameters of our model.

Following the definitions traditionally adopted in previous theoretical literature, we argue that the probabilities associated to a connection with a labeled node ($p_s^L$, $p_d^L$ and $p_l^{NL}$) quantify the relevance of endogenous mechanisms for link formation, given that the initiator of the alliance has information about the network position (i.e. social capital) of its potential partners. Likewise, the probabilities associated to a connection with a non-labeled node ($p_n^L$ and $p_{nl}^{NL}$) estimate the relevance of the exogenous mechanisms: in this case, the initiator cannot have any information about the social capital of a firm that is not part of the network yet. The five probabilities are bounded by two conditions, reducing the number of independent parameters to three; their nomenclature and their meaning are summarized in Table 2.

**Link formation.** After deciding the category of each of its $m$ partners, we assume that the initiator selects its specific partners within those categories according to their attractiveness. Indeed, it has been shown[19,20] that firms tend to connect to the firms having a higher prominence (i.e. history of previous alliances). We estimate this considering the degree of each potential partner. More precisely, we use a linear preferential attachment rule, where the probability to attach to a node $j$ linearly scales with its degree $k_j$, meaning that $\Pi(k_j) \sim k_j$. The preferential attachment rule is applied within the pool of all candidate partners, once the selection of the partner category has been made by the alliance initiator (see Fig. 6). This rule obviously does not apply when the initiator – be it labeled or not – decides to connect to a non-labeled node, which has by definition no previous partners ($k_j = 0$). In this case, the partner is selected among all non-labeled nodes with equal probability. When the selection process is complete, the initiator connects to its $m$ partners. In agreement with our representation of the R&D network, we assume that all the $m$ partners will also link to each other, forming a fully connected clique of size $m + 1$.

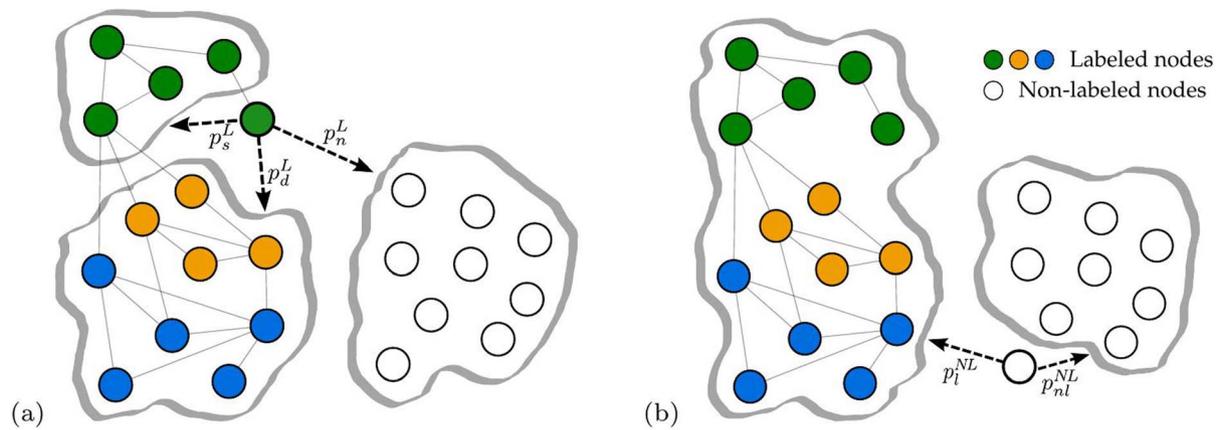

**Figure 6 | Selection of partner categories.** (a) If a labeled node (depicted in green) gets activated, it has 3 choices: it can link to a labeled node having the same label with probability $p_s^L$, or to a labeled node having a different label with probability $p_d^L$, or to a non-labeled node with probability $p_n^L$. (b) If a non-labeled node (depicted in white) gets activated, it has 2 choices: it can link to a labeled node with probability $p_l^{NL}$, or to another non-labeled node with probability $p_{nl}^{NL}$.

**Model implementation.** We perform extensive computer simulations by applying the above-described model and varying the values of its independent parameters. We fix the model parameters that we can directly measure from the data, namely the number of agents $N = 14,561$, the distribution of the node activities $a_i$, and the distribution of number of partners $m$ per alliance event. We stop every computer simulation when the total number of formed alliances equals the number of alliance events reported in the SDC dataset, $E = 14,829$.

We vary the values of $p_s^L$, $p_d^L$ and $p_{nl}^{NL}$ in discrete steps spaced by 0.05, in the interval (0, 1). The parameters $p_s^L$ and $p_d^L$ are bounded by the condition $p_n^L = 1 - p_s^L - p_d^L \geq 0$, meaning that their sum has to be smaller or equal to 1. This condition translates into 3,420 points to explore in the 3-dimensional parameter space, for each of which we run 200 simulations (for a total of 684,000 runs). Similarly to the empirical R&D network, we consider the final aggregated network resulting from each of the 684,000 computer simulations and we test it against the real data with respect to three properties: average degree $\langle k \rangle$, average path length $\langle l \rangle$ and global clustering coefficient $C$. We find that all such quantities are distributed around the empirical values, as we report in the SI in more detail. This testifies that our model well captures the topology of the observed network for a large set of free parameters. It should be noted that we have imposed a few features from the empirical network (number of nodes $N$ and alliances $E$, and the distributions of node activities $a_i$ and partners per alliance $m$). However, the distributions of the simulated $\langle k \rangle$, $\langle l \rangle$ and $C$ obtained by exploring the parameter space of the model, although centered around the real values, exhibit a fairly large variance (as reported in the SI). As a next step we aim at identifying which parameter combination is able to give the best match with the real R&D network. To this purpose, we use a Maximum likelihood approach. The peculiarity of this study is that, instead of having a set of observations against which we can validate our model, we only have one empirical point: the real R&D network. In particular, we cannot consider the three measures as independent, therefore the likelihood function $\mathcal{L}$ reads as:

$$\mathcal{L}(p|net^{OBS}) = f(net^{OBS}|p) \tag{3}$$

where $f(\cdot)$ is the joint density function of all parameter combinations $p$ resulting in a network that is equivalent to the observed one $net^{OBS}$. Both $p$ and $net^{OBS}$ are vectors with three components, expressing respectively the three model parameters $p \equiv (p_s^L, p_d^L, p_{nl}^{NL})$ and the three global network measures $net^{OBS} \equiv (\langle k \rangle^{OBS}, \langle l \rangle^{OBS}, C^{OBS})$. Therefore, we need to find the parameter combination $(p_s^L, p_d^L, p_{nl}^{NL})$ maximizing the likelihood $\mathcal{L}(p|net^{OBS})$ to generate a network whose macroscopic properties are *sufficiently similar* to the real network $net^{OBS}$. By this, we mean that the relative errors from the observed values for the average degree $\epsilon_{\langle k \rangle}$, the average path length $\epsilon_{\langle l \rangle}$ and the global clustering coefficient $\epsilon_C$ have to be smaller than a certain threshold $\epsilon^0$. We empirically compute the likelihood function $\mathcal{L}$ for each point in the parameter space by counting the fraction of its 200 simulation realizations that fulfill the criteria $\epsilon_{\langle k \rangle} < \epsilon^0$; $\epsilon_{\langle l \rangle} < \epsilon^0$; $\epsilon_C < \epsilon^0$. This way, we obtain values that can range from 0 (no realization of that parameter combination fulfills the criteria) to 1 (all of its realizations fulfill the criteria).

**Table 2 | Model parameters and their explanation.** We have two binding conditions, reducing the number of independent parameters to three: the probabilities $p_s^L$, $p_d^L$ and $p_n^L$ sum up to 1. Likewise, $p_{nl}^{NL}$ and $p_l^{NL}$ sum up to 1. We report the probabilities that we choose as independent parameters in bold character

| Parameter | Meaning | Type of mechanism |
|---|---|---|
| **$p_s^L$** | **Probability of a labeled node to select a node with the same label** | **Endogenous** |
| **$p_d^L$** | **Probability of a labeled node to select a node with a different label** | **Endogenous** |
| $p_n^L$ | Probability of a labeled node to select a non-labeled node | Exogenous |
| **$p_{nl}^{NL}$** | **Probability of a non-labeled node to select a non-labeled node** | **Exogenous** |
| $p_l^{NL}$ | Probability of a non-labeled node to select a labeled node | Endogenous |

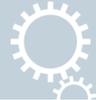

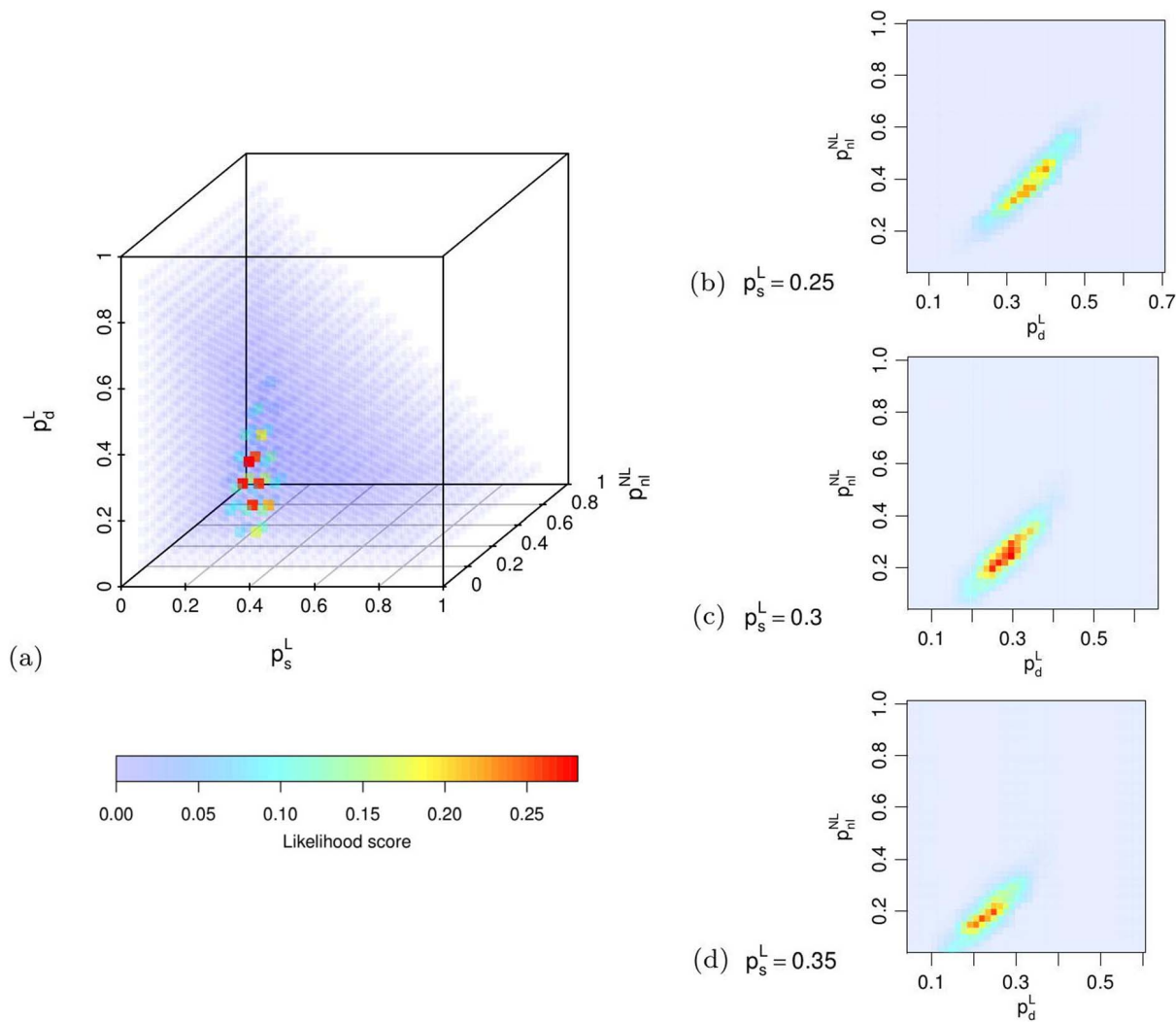

**Figure 7** | Likelihood scores for all points in the parameter space, for $\epsilon^0 = 2\%$, represented with a 3-dimensional color map (a). After fixing the value of $p_s^L$ to 0.25 (b), 0.3 (c) and 0.35 (d), we report the likelihood score as a function of $p_d^L$ and $p_{nl}^{NL}$, using the same color scale.

The error threshold value $\epsilon^0$ we impose for the computation of the likelihood score influences the number of points in the parameter space that fulfill our matching criteria. Obviously, by decreasing $\epsilon^0$, we observe a smaller number of points displaying high likelihood scores, as we could expect, because a better representation of reality is required (see SI). We take a conservative approach and use an error threshold $\epsilon^0 = 0.02$, that ensures a good matching with the observed R&D network, without cutting out too many points in the parameters space. The corresponding Likelihood scores are reported in Fig. 7 by means of a 3-dimensional color map, where the color scale is representative of the likelihood. To have a more detailed representation of the likelihood scores, we also show three slices of the parameter space obtained by fixing the parameter $p_s^L$ in the range $0.25 \div 0.35$, corresponding to the highest likelihood score region, always using the error threshold $\epsilon^0 = 0.02$. The 2-dimensional color maps reported in Fig. 7 depict the likelihood score as a function of the other two free parameters $p_d^L$ and $p_{nl}^{NL}$.

**Table 3** | Model parameter set $p^*$ defining the optimal simulated R&D network. The average degree, average path length and global clustering coefficient of the 200 realizations of the optimal R&D network are compared to their analogous empirical values

| Optimal simulated R&D network | | | | Observed R&D network | |
|---|---|---|---|---|---|
| Model parameter | Value | Measure | Value | Measure | Value |
| $p_s^{*L}$ | 0.3 | $\langle k \rangle^*$ | $2.764 \pm 0.006$ | $\langle k \rangle^{OBS}$ | 2.736 |
| $p_d^{*L}$ | 0.3 | $\langle l \rangle^*$ | $5.329 \pm 0.068$ | $\langle l \rangle^{OBS}$ | 5.412 |
| $p_n^{*L}$ | 0.4 | $C^*$ | $0.098 \pm 0.005$ | $C^{OBS}$ | 0.101 |
| $p_{nl}^{*NL}$ | 0.25 | | | | |
| $p_l^{*NL}$ | 0.75 | | | | |

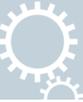

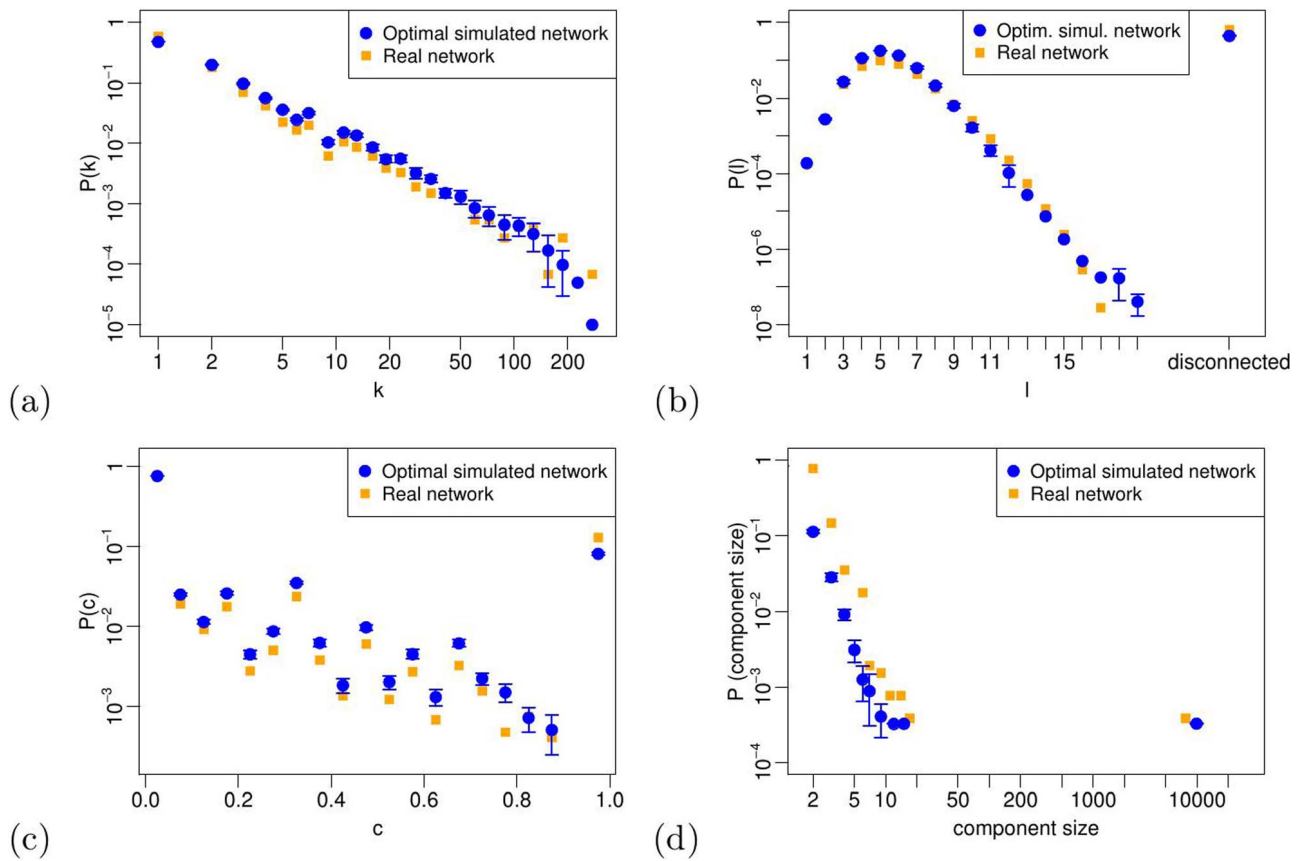

**Figure 8 | Distributions of node degrees (a), path lengths (b), local clustering coefficients (c) and component sizes (d) for the real and the optimal simulated networks.** Most of the error bars are not visible, because the values are very narrowly distributed across the 200 realizations of the optimal simulated network.

The point with the highest likelihood score has the following coordinates in the parameter space: $p_s^{*L} = 0.3$, $p_d^{*L} = 0.3$ and $p_{nl}^{*NL} = 0.25$. We can already see that in the optimal configuration, labeled nodes exhibit a balanced alliance strategy, with $p_s^{*L} = 0.3$, $p_d^{*L} = 0.3$, and consequently $p_n^{*L} = 0.4$, while the non-labeled nodes exhibit a strong tendency to connect to labeled nodes ($p_l^{*NL} = 0.75$), as opposed to a low linking probability with other non-labeled nodes ($p_{nl}^{*NL} = 0.25$). We report in Table 3 the set of parameter values maximizing the likelihood score, together with the values of average degree, average path length and global clustering coefficient for the simulated and the real R&D networks. From now on, we call the network generated with these parameters the *optimal simulated R&D network*. More precisely, we generate 200 realizations of the optimal simulated R&D network (as well as of any other network with a generic parameter set). For this reason, the results we present in the next section are computed on all the 200 realizations of such optimal network.

The optimal set of linking probabilities gives some interesting insights into the nature of the strategies pursued by firms when forming R&D partnerships. Indeed, all firms tend to have a preference to link incumbent firms: 60% of the alliances initiated by incumbents belong to this category ($p_s^{*L} + p_d^{*L}$), as well as 75% of the alliances initiated by newcomers $p_l^{*NL}$. This result is in line with well-known economic theories[33] that have shown how previous interactions between two firms increase the likelihood of future alliances among them if they are already part of the R&D network. In addition, newcomers are incentivized to join the R&D network by partnering with firms that are already part of it[34]. On the other hand, we find that 40% of the alliances initiated by incumbents, as well as 25% of the alliances initiated by newcomers, are directed to new-comers. These alliances can be driven only by exogenous factors[17], and a possible explanation behind this tendency is the appealing of newcomers' commercial or technological capital.

Overall, our findings suggest that both endogenous and exogenous mechanisms contribute to the alliance formation. However, the first class appears to be more prominent: the fine tuning of our model provides additional evidence, and a precise quantification, of how previous network structures play the biggest role in deciding the potential partners when a new alliance is formed. As reported in the literature[20,33], the belonging to the R&D network, and in particular the belonging to a specific circle of influence, signals a firm's reliability and competencies to potential partners. This mechanism is clearly predominant over the exogenous search for alliance partners, hence we argue that being aware of the partners' positions in the R&D network is of fundamental importance for every firm.

**Model validation.** The optimal simulated R&D network, as we have shown above, is generated by the set of parameter values $p^* \equiv (p_s^{*L} = 0.3; p_d^{*L} = 0.3; p_n^{*L} = 0.4; p_{nl}^{*NL} = 0.25; p_l^{*NL} = 0.75)$. We now want to check whether our model, fed with this optimal parameter set, is able to reproduce further microscopic properties of the real network. To this purpose, we report in Fig. 8 four additional distributions computed on the optimal simulated R&D network – node degrees, path lengths, local clustering coefficients and component sizes – and compare them to the empirical ones (see Fig. 3). From now on, in every plot we show, the blue circles correspond to the mean values and the error bars correspond to the standard deviations of all the quantities we analyze on the 200 realizations of the optimal simulated R&D network.

Remarkably, we find that our model is able to reproduce all the distributions, namely the typical right-skewed degree distribution,

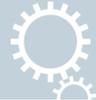

the path length distribution peaked around the mean value 5 and the local clustering coefficient distribution. The model can also reproduce the component size distribution, showing the emergence of a giant component in the network (containing roughly 60% of the nodes) together with many smaller components down to size two. Isolated nodes (nodes with degree equal to 0) are excluded from our representation; hence, the smallest observable component size in our networks is 2.

Going further in our validation procedure, we test the modular properties of the optimal simulated R&D network, by running the Infomap community detection algorithm[30] on all of its realizations. We identify the presence of $1,600 \pm 20$ clusters (whilst 3,500 clusters populate the empirical R&D network), whose minimum size is 2 and maximum size is around 100 nodes, similarly to the empirical network (see Fig. 4). We report in Fig. 9 a visual representation of the optimal simulated R&D network and the size distribution of the detected clusters. Interestingly, this distribution resembles the one of the empirical R&D network, with the only exception of having significantly fewer counts related to small clusters of size 2 and 3. The larger clusters, up to 100 nodes, that dominate the network structure and contribute to its modularity, are equally populating the empirical and the optimal simulated R&D networks. Another evidence of their similarity is the modularity score of the optimal simulated R&D network $Q^* = 0.66 \pm 0.01$, surprisingly close to its empirical analogue $Q^{OBS} = 0.68$. Also in the case of the optimal simulated R&D network, its modularity score $Q^*$ is significantly greater (with a $p$-value computationally indistinguishable from zero) than the ones obtained for a set of 500 randomly generated networks with the same degree sequence (whose $Q$ is normally distributed around 0.485 with a standard deviation of 0.001), showing that the modularity is not an artifact of the network size and density.

We now test whether our node labels are actually able to reproduce such a modular structure of the network. In order to estimate the overlap between the clusters detected via the Infomap algorithm and the circles of influence defined by our node labels, we compute the *normalized mutual information* coefficient $I_{norm}$[35], very often used to this purpose[36]. Given two network partitions $A$ and $B$, the value of the coefficient $I_{norm}(A, B)$ ranges from 0 (if the partitions $A$ and $B$ are independent) to 1 (if the partitions $A$ and $B$ are identical). In our case, we obtain a striking $I_{norm}$(Labels, Infomap clusters) $= 0.899 \pm 0.001$, testifying how well our node labels capture the emergence of clusters in the R&D network. This result is even more remarkable if we think that the Infomap algorithm detects clusters based on the probability flow of random walks in the network[30], while our label propagation mechanism only consists in an assignment of a fixed membership attribute. We also present a visual comparison of the clusters identified by means of Infomap with the circles of influence resulting from the implementation of our model in Fig. 9. Similarly to the empirical R&D network, we consider only the 30 largest Infomap clusters in the optimal simulated R&D network and visualize them by grouping the corresponding nodes in distinct regions of the plot; in addition, here we depict our node labels with arbitrary colors. As testified by the high normalized mutual information score, our visual example nicely confirms that most of the nodes in a given cluster share the same label. The size distribution of the circles of influence defined by these labels is also shown in Fig. 9. Its similarity to the size distribution of the Infomap clusters in both the empirical and the optimal simulated R&D network provides another evidence of the goodness of our model.

In order to estimate to what extent our link formation rules capture the decision making process made by real firms, we test the optimal simulated network with respect to a feature that is both *microscopic* and *dynamic*: the distribution of path lengths between every pair of nodes at the moment of the link formation. This should not be confused with the path lengths analyzed before, whose distribution was computed on the final aggregated R&D network, between *every* pair of nodes, in both the real and the simulated case. Now we only consider pair of nodes that eventually form a link between each other. More precisely, we plot the distribution of the path lengths between two firms as of the day before their alliance formation (for the real R&D network) and the path lengths between two nodes at the time step preceding the link formation (for the optimal simulated R&D network). We also consider as separated counts all those alliance events involving at least one newcomer firm (or an isolated node, in the simulated network).

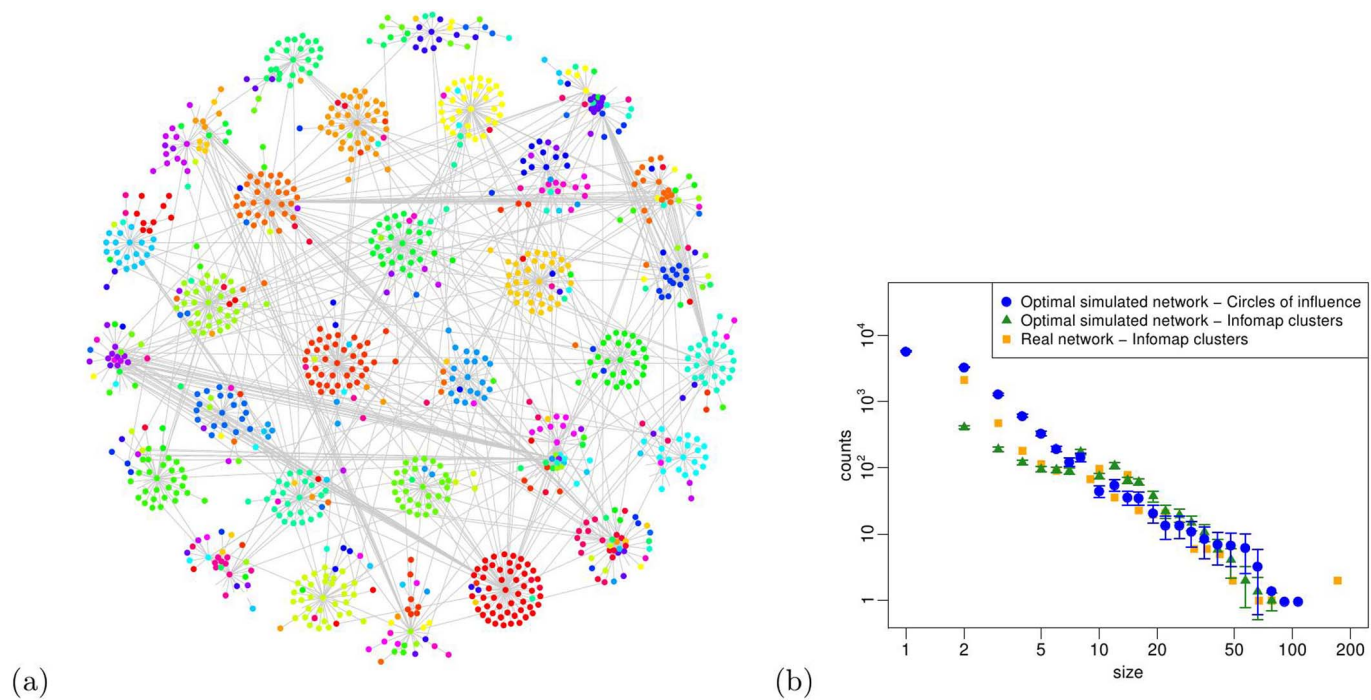

(a)                                                    (b)

**Figure 9** | (a) Visual representation of one realization of the optimal simulated R&D network, using the Fruchterman-Reingold layout[42] and considering only the 30 largest clusters detected by the Infomap algorithm. Distinct clusters are represented by grouping nodes in distinct regions of the plot area. In addition, we depict our node labels by using different colors; it is clearly observable that most of the nodes in a given cluster share the same label. (b) Size distribution of i. the circles of influence in the 200 realizations of the optimal simulated R&D network, ii. the Infomap clusters in the 200 realizations of the optimal simulated R&D network and iii. the Infomap clusters in the empirical R&D network.

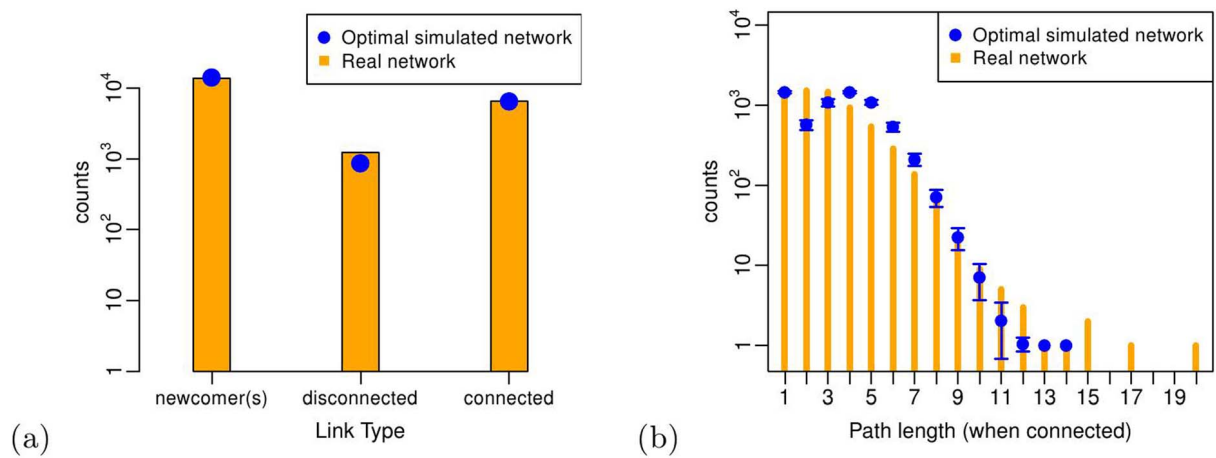

**Figure 10** | (a) Distribution of link types for the real and the simulated R&D networks. "Connected" refers to nodes already belonging to the same connected component of the network prior to the link formation; "disconnected" refers to nodes already belonging to the network, but placed in two disconnected components; "newcomer(s)" means that at least one of the nodes was isolated (i.e. not yet part of the network) before the link formation. (b) Distribution of path lengths at the moment of link formation (only for nodes belonging to the same connected component).

We show our findings in Fig. 10. The model can reproduce the counts of links formed between (i) firms belonging to the same connected component of the network, (ii) firms belonging to different disconnected components and (iii) involving at least one newcomer (isolated) firm. Furthermore, the model reproduces also the counts relative to nodes which are already connected by a path before the link formation. The only small discrepancies can be observed in correspondence to path lengths equal to 2 and 3, due to effects of triadic and cyclic closure exhibited by real firms that are not fully captured by our model, as we already anticipated. However, our model correctly predicts the formation of links between nodes that are relatively distant in the network or even already directly connected – the cases when the path length is equal to 1 are related to the same two partners engaging in a new alliance.

In conclusion, we find that our model, although tuned only considering three global static measures, provides a surprisingly good prediction of several microscopic and dynamic features, such as the distributions of degree, local clustering, path length and component size, the emergence of network clusters and, even more remarkably, the distribution of path lengths at the moment of the alliance formation.

## Discussion

In the present work, we introduced an agent based model aimed at reproducing the formation of a collaboration network, namely a global inter-organizational R&D network. The agents, representing real firms, are endowed with two key attributes: an activity (representing their propensity to engage in new alliances) and a label (representing their membership in a given circle of influence). We extended the concept of activity from the field of temporal networks to economic networks, by empirically measuring it for the first time on a set of real firms and subsequently implementing it into an agent based model. Next, we proposed a simple yet effective set of rules to reproduce the topology of the observed R&D network. Our model is centered around the assumption that the firms have a membership attribute, that we call *label*. Such attribute can be propagated to other firms as a consequence of an alliance, thus defining the so called *circles of influence* (groups of firms sharing the same membership attribute). The model includes different link formation probabilities, that depend on both the alliance initiator's and its future partners' membership attributes. By running extensive computer simulations, and imposing only a few features from the empirical network (number of nodes and alliances, distributions of node activities and partners per alliance), we generated a set of networks that we compared to the observed R&D network, with respect to three global properties: average degree, global clustering coefficient and average path length. The distributions of such quantities across the entire parameter space, although centered around the empirical values, show a fairly large variance (see SI). Through a Maximum likelihood approach, we then identified the set of linking probabilities (i.e. the model free parameters) that generates the closest network to the observed R&D network, obtaining the following result. As summarized in Table 3, when the initiator of the alliance is a labeled node (i.e. an incumbent firm), it connects to a node having the same label with probability $p_s^L = 0.3$, to a node having a different label with probability $p_d^L = 0.3$ and, consequently, to a non-labeled node (i.e. a newcomer firm) with probability $p_n^L = 0.4$. When the alliance is initiated by a non-labeled node (a newcomer), it connects to a labeled node with probability $p_l^{*NL} = 0.75$ or to another non-labeled node with probability $p_{nl}^{*NL} = 0.25$. The optimal simulated network generated by our model exhibits values of average degree, global clustering coefficient and path length that deviate from the empirical values by less than 2%.

The linking probabilities we listed above have a precise meaning in terms of strategies pursued by the firms willing to form R&D partnerships. Our findings suggest that incumbent firms tend to have a preference towards other incumbent firms: 60% of their alliances belong to this category, split between a 30% probability to connect to a node in the same circle of influence and a 30% probability to connect to a node in a different circle of influence. This finding is in agreement with well-known economic theories[33,34] pointing out that previous network connections positively affect the likelihood of alliance formation between two companies. Moreover, we extend previous empirical results[9] by including an explicit quantification of the linking probabilities. We find that incumbents willing to form alliances with other incumbents equally share their preference between firms belonging to the same circle of influence and firms belonging to a different one. In the remaining 40% of the cases, incumbents form alliances with newcomers: these alliances are driven only by exogenous factors[17], since there cannot be any network endogeneity affecting nodes that are not part of the network yet.

On the other hand, newcomers have an even more unbalanced alliance strategy, given that they link to incumbent firms in 75% of the cases. Such alliances are driven by network endogenous factors, namely the newcomers' motivation to join the R&D network by partnering with firms that are already part of it. This is in line with a number of studies[20,37] that analyzed how being embedded in the network signals attractiveness, also beyond the firm's circle of influence and even to newcomer firms. Indeed, the preferred way for the newcomers to enter the R&D network is to form an alliance with an incumbent firm. Our results confirm and extend previous findings[9,38] that did never quantify such a preference of newcomers towards incumbents. However, a fraction (25%) of alliances initiated by newcomers are directed to other newcomers. The reasons behind these alliances are not related to network endogeneity, but rather to exogenous factors such as the firms' commercial or technological capital. Some newcomers prefer to join the R&D network by partnering with other newcomers with no network experience[39] – this could be the case, for instance, of small *Pharma* or *IT* companies – rather than engaging in an alliance with an incumbent firm.

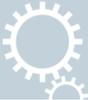

Overall, the fine tuning of our model suggests that endogenous mechanisms for network formation are predominant over the exogenous ones, as testified by the linking probabilities towards firms that are already part of the R&D network. We find that incumbents are the preferred alliance partners for both other incumbents (60%) and newcomers (75%), thus providing new evidence and quantification of how relevant existing network structures are in selecting partners when new R&D alliances are formed.

We further validated our model by testing microscopic network properties. Without imposing any equivalence criterion on these quantities, we obtained a surprising agreement with the empirical data. Our agent based model, fed with the optimal parameter combination, is able to reproduce the distributions of degrees, path lengths, local clustering coefficients and network component sizes. Remarkably, we also reproduced the emergence of clusters in the R&D network, that can be identified with our circles of influence. Precisely, we found a 90% overlap between the network partition defined by our labels and the one we detected by means of a very used community detection algorithm (Infomap). We argue that this highlights how effectively the label propagation mechanism can model the expansion of the firm circles of influence within the R&D network. Finally, we reproduced the distribution of path lengths between every pair of nodes at the moment of link formation, clearly indicating the goodness of our model at capturing the microscopic rules driving the alliance formation between real firms.

Although the model captures many features of the empirical R&D network, it can be further improved to account for other real world observations. One of the limitations is assuming fixed node labels; this condition could be relaxed by introducing a label decay, representing the exit of a firm from its circle of influence. Such an extension might be useful especially when validating a dataset with a longer time extension. A second limitation is that the alliance partners chosen by the initiators have no power in accepting this invitation; such a realistic attachment rule could be included in the model, at the price of requiring more parameters. In addition, a linking preference towards partners of partners could be added to the model, to better reproduce the observed effects of triadic closure in the R&D network[40,41]. A last possible extension is represented by a more rigorous definition of the exogeneity rules, resulting in a quantification of the effects of the firms' technological and commercial capital.

Finally, we believe that our model can reproduce the topology of other collaboration networks, possibly with a different optimal parameter set. Similarities and differences in the numerical values of the linking probabilities could provide insights into the microscopic rules driving the alliance formation in different collaboration networks.

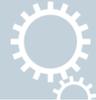

## Acknowledgments

We thank Alessandro Vespignani, David Garcia, René Pfitzner, Ingo Scholtes and Rebekka Burkholz for useful comments and discussions. MVT acknowledges financial support from the SNF through grant 100014 126865. FS acknowledges support by the EU-FET project MULTIPLEX 317532.

## Author contributions

M.V.T., N.P., C.J.T., M.K. and F.S. designed the research and participated in the writing of the manuscript. M.V.T. analysed the empirical data and performed the numerical calculations.

## Additional information

**Supplementary information** accompanies this paper at http://www.nature.com/scientificreports

**Competing financial interests:** The authors declare no competing financial interests.

**How to cite this article:** Tomasello, M.V., Perra, N., Tessone, C.J., Karsai, M. & Schweitzer, F. The role of endogenous and exogenous mechanisms in the formation of R&D networks. *Sci. Rep.* **4**, 5679; DOI:10.1038/srep05679 (2014).